\documentclass[secnumarabi,superscriptaddress,amssymb,nobibnotes,aps,prc]{revtex4}

\usepackage{epsfig}
\usepackage{graphicx}
\usepackage{amsmath}
\usepackage{hyperref}

\begin{document} 
\title{Mapping color fluctuations in the photon in ultraperipheral  heavy ion collisions at the Large Hadron 
 Collider}
  
\author{M. Alvioli}
\affiliation{Consiglio Nazionale delle Ricerche, Istituto di Ricerca per la 
  Protezione Idrogeologica, via Madonna Alta 126, I-06128 Perugia, Italy}
  
\author{L. Frankfurt}
\affiliation{Particle Physics Department, School of Physics and Astronomy, Tel Aviv
  University, 69978 Tel Aviv, Israel}
\affiliation{Department of Physics, the Pennsylvania State University, State College, PA 16802, USA}

\author{V. Guzey}
\affiliation{National Research Center ``Kurchatov Institute'', Petersburg Nuclear Physics Institute (PNPI), Gatchina, 188300, Russia}

\author{M. Strikman}
\affiliation{Department of Physics, the Pennsylvania State University, State College, PA 16802, USA}

\author{M. Zhalov}
\affiliation{National Research Center ``Kurchatov Institute'', Petersburg Nuclear Physics Institute (PNPI), Gatchina, 188300, Russia}

\pacs{24.85.+p, 25.20.Lj, 12.40.Vv}

\begin{abstract} 
We model  effects of color fluctuations (CFs) in the light-cone photon  wave function  and for the 
first time make predictions for the distribution over the number  of wounded nucleons $\nu$ in the inelastic photon--nucleus 
scattering.  We show that CFs lead to a dramatic enhancement of this distribution at $\nu=1$ and large $\nu > 10$. We also 
study the implications of  different scales and CFs in the photon wave function on the total transverse energy $\Sigma E_T$ 
and other observables in inelastic $\gamma A$ scattering with different triggers. 
Our predictions can be tested in proton--nucleus and nucleus--nucleus ultraperipheral collisions 
at the LHC and will help to map CFs, whose first indications have already been observed at the LHC. 
\end{abstract}

\maketitle 

\section{Introduction}
\label{sec:Intro}
One of key features of high energy processes in the target rest frame is that the wave function of a projectile is the 
superposition of coherent (so-called frozen) configurations~\cite{Gribov:1968ia,Gribov:1973jg}, 
which is a consequence of the  uncertainty principle and Lorentz slowing down of the  interaction time.
In the pre-QCD  times,  coherence of high energy processes  
has  been   extensively studied   in the photon--nucleon ($\gamma N$) 
and photon--nucleus ($\gamma A$) collisions, for a review, see~\cite{Bauer:1977iq}.  
In particular, it was established that the resolved photon is dominated by the contribution of the light vector meson 
component of the photon wave function, which is responsible for about 70\% of $\sigma_{\rm tot}(\gamma N)$. 
The origin of the photon components, which are responsible for the 
 remaining 30\% of the $\gamma N$ cross section,  is a matter of debate. 

In QCD coherence of high energy processes is  well understood theoretically and  established experimentally, 
for a review, see, e.g.~\cite{Frankfurt:1994hf,Frankfurt:2000tya}.
A distinctive feature of the  QCD dynamics is that the interaction strength of different configurations 
of quarks and gluons, which are QCD constituents of projectile hadrons, photons, etc., varies.  We refer to this 
phenomenon as color fluctuations (CFs).  In the literature one alternatively uses the term cross section fluctuations, 
which refer predominantly to soft hadron (photon) interactions at high energies. 
  
A particular dramatic example of CFs is the phenomenon of color transparency (CT) when,
as a consequence of color screening,
the strength of the interaction of a high energy hadron (photon)   
in a configuration with  a small transverse size is much  smaller than the average interaction strength,  
for a recent review, see~\cite{Dutta:2012ii}.  While CT is a natural mechanism for the
interaction with the strength smaller than the average one, several mechanisms like fluctuations of transverse 
size,  gluon density, the phenomenon of spontaneously broken chiral symmetry,
  etc.~can contribute to fluctuations with the larger-than-average interaction strength.  

It has been demonstrated  long ago  by the direct calculations that  the contribution  of planar diagrams 
to the  total cross section of a hadron--hadron collision tends to zero with an increase of the collision energy~\cite{Mandelstam:1963cw}.  Therefore,  the contribution of  consecutive multiple  rescatterings  
of a  hadron projectile to the total cross section of the hadron--nucleus scattering described by  planar Feynman diagrams  
 rapidly decreases with an increase of the invariant collision energy $s$~\cite{Gribov:1968jf}.  
 Thus,  within a quantum field theory 
 multiple interactions of the projectile are dominated  at high energies by the contribution of non-planar diagrams.  
The  Gribov--Glauber approximation~\cite{Gribov:1968ia} has been suggested to  resolve this theoretical puzzle. 
 It   accounts for the contribution of non-planar diagrams  and  employs 
 duality between  non-planar diagrams  and a sum of the elastic contribution 
and the diffractive intermediate states (duality between  $s$ and $t$ channels)  to rewrite  
formulae in the form rather similar to the Glauber approximation~\cite{Gribov:1968ia}. 
This  theoretical description accounts for coherence of high energy processes and predicts that the geometry of  
$hA$ collisions  should be rather close  to that  expected  within the Glauber approach.  Hence the 
Gribov--Glauber approximation is routinely used in the evaluation of geometry of the heavy ion collisions.
By virtue of duality the Gribov--Glauber approximation includes diffractive intermediate states,  which allow one 
to account for energy--momentum conservation, see the discussion in~\cite{Alvioli:2014sba}. 
The presence in the formulae of  the contribution of  inelastic diffractive states leads to the inelastic shadowing 
correction for $\sigma_{tot}(pA)$~\cite{Gribov:1968jf}.  The inelastic shadowing correction was evaluated in a number 
of papers and found to agree well with the data, see discussion and references in, e.g.,~\cite{Alberi:1981af}.  

It has been suggested that the interaction matrix  of the initial hadron or diffractively produced hadronic states 
with target nucleons, which arises within Gribov--Glauber approach, can be diagonalized~\cite{Good:1960ba,Miettinen:1978jb}. 
In the particular case,  when  diffractive intermediate states are  resonances,  this diagonalization 
has been performed in~\cite{Frankfurt:1997zk}.
The method of CFs developed in~\cite{Baym:1995cz} and discussed below is the further generalization of the Gribov--Glauber 
approximation, which  allows one to account  for the fluctuations of the interaction strength 
and other implications of QCD.

 Several effects were observed at  collider energies,  
which naturally emerge in the CF framework.  First, the 
ATLAS study~\cite{Aad:2015zza} of the charged-particle pseudorapidity distribution $dN_{\rm ch}/d\eta$ 
in proton--lead ($pPb$) collisions at $\sqrt{s_{NN}}=5.02$ TeV
as a function of centrality and the pseudorapidity $\eta$ showed that CFs affect the collision geometry 
by broadening the distribution of the number of 
participating nucleons $N_{\rm part}$ for large $N_{\rm part}$ (Fig.~13 of Ref.~\cite{Aad:2015zza}). It
can be interpreted as broadening of the distribution in the number of wounded nucleons 
naturally emerging in the CF approach~\cite{Baym:1995cz} 
(note that in these early papers, CFs were called cross section fluctuations).  It
results in a milder dependence of $dN_{\rm ch}/d\eta(\langle N_{\rm part}\rangle/2)$ on centrality, 
especially at large negative rapidities in the Pb-going direction
 (Figs.~11 and 12 of Ref.~\cite{Aad:2015zza}) than that expected in the  combinatorics of the geometric 
 Gribov--Glauber model~\cite{Bertocchi:1976bq}. The numerical results for the distribution of CFs 
 in the proton used in the analysis of~\cite{Aad:2015zza} are consistent with the  expectations of~\cite{Guzey:2005tk,Alvioli:2013vk}.

The second effect is the observation of a  large violation of the Gribov--Glauber approximation for the dependence 
of the jet production on the centrality observed in $pA$ collisions at the LHC~\cite{ATLAS:2014cpa} 
and in $dA$ collisions at RHIC~\cite{Adare:2015gla},  for which a large-$x$  parton 
momentum fraction of  the proton is involved. The central-to-peripheral $R_{\rm CP}$ ratio is suppressed by as 
much as 80\% at the LHC and 50\% at RHIC at the largest measured $p_T$.  At the same time the combinatorics given 
by the geometric Glauber  picture works very well for the collisions with up to eight nucleons, if $ x $ is small enough, $
x\le 0.1$. While CFs only increase the deviation of $R_{\rm CP}$ from unity,  this pattern is consistent with the 
$x$-dependence of  CFs  expected within QCD~\cite{Alvioli:2014eda}.

The third effect is the 
significant suppression  of the rate of $\rho$ meson production in the coherent 
$\gamma A\to \rho A$ reaction measured in Pb-Pb ultraperipheral collisions (UPCs)
at the LHC~\cite{Adam:2015gsa} as compared to the expectations of the 
vector dominance model combined with the Gribov--Glauber approximation 
for the photon--nucleus interaction.  This was explained in~\cite{Frankfurt:2015cwa} by taking into account the effect 
of CFs in the photon wave function, which reduce the effective $\rho$--nucleon cross section by suppressing the  overlap 
of the vector meson and photon wave functions and lead to sizable inelastic (Gribov) nuclear shadowing due to 
the photon inelastic diffraction into large masses.

In this paper we argue that one can  map CFs in the photon wave function using  
ultraperipheral collisions (UPCs) of heavy ions at the LHC. Although feasibility of UPC studies was analyzed at length 
in~\cite{Baltz:2007kq}, the studies discussed below were not addressed. 
Primarily this is because such analyses  became feasible due to  the experience accumulated in the analysis 
of $pA$ collisions at the LHC. In a long run studies along these lines at the Electron-Ion 
Collider~\cite{Accardi:2012qut,Boer:2011fh} would provide a detailed information 
on CFs in the photon and their dependence on the photon virtuality. 
The main challenge for building a realistic description  of the photon--nucleon (nucleus) interactions  at collider 
energies  is to take into account  the multi-scale  structure of the light-cone wave function of the photon associated with 
presence of soft and hard intrinsic scales.  In particular, the photon wave function contains  several  types of
configurations: large-$\sigma$ configurations characterized by small transverse momenta $k_t < 0.5$
GeV and invariant masses comparable to the masses of light vector mesons interacting
with the strength $\sim  \sigma_{\pi N}$ ($\sigma_{\pi N}$ is the total pion--nucleon cross section), 
configurations interacting with $\sigma$  much
larger than $\sigma_{\pi N}$ related to the presence of soft large-mass diffraction, and
small-$\sigma$ configurations with large $k_t \geq 1$ GeV, whose contribution   results  in the  leading twist nuclear shadowing.
UPCs at the LHC  correspond to  a wide  interval of invariant energies $W \le 500$ GeV, where hard physics should be 
well  described within the DGLAP approximation, see \cite{Altarelli:2001ji}  and references therein.   Thus,  
a  more rapid increase of parton  distributions with energy at extremely small $x$, which
is often discussed in the literature,  is beyond the scope of this paper.
 
We propose a model of CFs in the hadronic component of the
photon  wave function by combining the information obtained in the 
analysis of photoproduction  of $\rho$ mesons at the LHC energies~\cite{Frankfurt:2015cwa},
which enables us to model the photon configurations interacting mostly with the strength exceeding the typical
$\rho$--nucleon cross section,  with that obtained in photoproduction of $J/\psi$ mesons~\cite{Guzey:2013jaa},
which is amenable to the perturbative QCD (pQCD) description of the weakly-interacting configurations. 
Since the  information on coherent photoproduction of $\rho$ and  $J/\psi$ is available for $ W\sim 100$ GeV, we focus in this paper 
on this energy range.  The $W$ dependence of the discussed effects for higher $W$ will be considered elsewhere.

We apply the resulting model of the photon CFs to the calculation of the distribution over the number of 
 wounded nucleons, $\nu$, involved  in the inelastic $\gamma A$ scattering.
 We show that as a consequence of CFs
around the average value, the soft inelastic nuclear shadowing effect 
is strongly enhanced as compared to $pA$ collisions. We also take into account an additional effect 
of the different pattern of the interaction of small dipoles, which leads to the leading twist nuclear shadowing 
and which is absent in the Gribov--Glauber approximation.  This effect leads to the significant probability for small 
dipoles  to interact with several nucleons,
which noticeably reduces the distribution over $\nu$ for small values of $\nu$.

This paper is organized as follows. In Sect.~\ref{sec:photon} we develop a 
model for CFs in the photon wave function for the photon--nucleon interaction.
In Sect.~\ref{sec:fluct} we present and discuss our predictions for the distribution over the number of wounded nucleons 
(inelastic interactions)  in the inelastic photon--nucleus scattering. In the 
calculations we use our model for CFs in the photon without and with 
an additional effect of the leading twist nuclear shadowing for the configurations interacting with small cross sections.
In Sect.~\ref{sec:transverse_energy},
we make a prediction for the transverse energy $\sum E_T$ distribution in $\gamma A$ collisions
using as a starting point
 the model of~\cite{Aad:2015zza} for the dependence of $\sum E_T$ on $\nu$. Finally, in Sect.~\ref{sec:conclusions} we 
discuss possibilities of special triggers, which would allow one  to use $\gamma A$ scattering  
to map out different components of the photon wave function.

\section{Color fluctuations in $\gamma A$ scattering: general formalism}
\label{sec:photon}

At sufficiently high photon energies $E_{\gamma}$ in the target rest frame, 
the coherence length associated with the hadronic fluctuation (component) of the photon wave function
of mass $M$ exceeds  the target radius $R_T$, $l_{\rm coh}=2E_{\gamma}/M^2 > R_T$.
In this case, the forward photon--target amplitude (the total photoabsorption cross section) 
can be expressed in terms of the dispersion representation over the masses  $M^2$~\cite{Gribov:1968gs}:
\begin{equation}\sigma_{\gamma N} =
{\alpha_{\rm e.m.}\over 24 \pi^2}  \int {dM^2\over M^2}
R_{e^+e^-\to {\rm hadrons}} (M^2) \sigma_{M N} \,,
\label{disp}
\end{equation} 
where $\alpha_{\rm e.m.}$ is the fine structure constant;  $R_{e^+e^-\to {\rm hadrons}} (M^2)=
\sigma(e^+e^-\to {\rm hadrons})/\sigma(e^+e^- \to \mu^+\mu^-)$ is the ratio of the 
$e^{+}e^{-}$ annihilation cross sections into hadrons (everything) and a muon pair, respectively, of a given 
invariant mass squared $M^2$; $\sigma_{M N}$ is the total cross section for the interaction of a given component 
with the target. It is important to emphasize that non-diagonal transitions between different photon components have 
been neglected  in Eq.~(\ref{disp}), which can be justified at present in the case of a heavy nuclear target~\cite{Gribov:1968gs}.

In the vector meson dominance model (VMD), 70\% of  the integral in Eq.~(\ref{disp})  is due to the sum of  
$\rho$, $\omega$ and $\phi$ mesons, which interact with hadrons with a strength similar to that of a pion (for $\rho$ 
and $\omega$)~\cite{Sakurai:1960ju,Feynman:1973xc,Bauer:1977iq}. 

A straightforward generalization of Eq.~(\ref{disp}) to the case of deep inelastic scattering (DIS) leads to a gross violation of the 
approximate Bjorken scaling, and hence to the contradiction with the leading twist QCD expectations. In the framework of the parton 
model, the qualitative
resolution of the  paradox was suggested by Bjorken~\cite{Bjorken:1972uk} by assuming that the interaction is dominated by the 
so-called aligned quark--antiquark pairs, where the quarks share asymmetrically the photon longitudinal momentum and have 
small transverse momenta $p_t$.  Such aligned quark--antiquark pairs configurations are strongly interacting with a nuclear target and 
correspond to typical vector meson-like (and/or other hadronic) configurations. Their contribution to the total virtual photon--nucleon cross section 
$\sigma_{\gamma^{\ast}N}$  is suppressed by a factor of $\mu^2/M^2$, where $\mu$ is a soft QCD scale, which leads to the scaling of 
$\sigma_{\gamma^{\ast}N}$.  

In QCD the  situation is somewhat different~\cite{Frankfurt:1988nt}: 
in addition to the aligned pairs,  configurations with large $p_t$ 
also contribute to $\sigma_{\gamma^{\ast}N}$;  their noticeable contribution is proportional to 
$\alpha_s(p_t^2)/p_t^2$, where $\alpha_s$ is the strong coupling constant,
and grows with an increase of the collision energy.

 Overall this leads to  the following approximate picture of the 
 hadronic component of the wave function of 
 the photon:   the majority  of the  configurations interact with strengths similar to the one given by
 CFs  in the $\gamma \to \rho,\omega $ transitions; they dominate at large and medium  
 $\sigma \ge \sigma_{\pi N}$. (They also include the fluctuations in the aligned jet component.)
 Note that with an increase of collision energies, these configurations 
  are likely to be somewhat  more localized than those in the elastic vector meson--nucleon  scattering~\cite{Frankfurt:2015cwa}.  
  In addition, there is a component which dominates for small $\sigma $ and which is described by the 
 perturbative (dipole)  wave function interacting with the strength given by perturbative QCD.

The formalism of cross section fluctuations was introduced before advent of QCD to explain presence of inelastic diffraction 
at small $t$~\cite{Good:1960ba, Miettinen:1978jb}. Its connection to the Gribov inelastic shadowing for double 
scattering was pointed out in \cite{Kopeliovich:1978qz}.
The basic idea of this approach is to diagonalize  the interaction matrix which arises in the Gribov--Glauber approach 
in the basis of elastic and diffractive states. The obtained  matrix describes the distribution over the values of the
cross section.    If diffractive states 
are hadron resonances, this program can be effectively performed~\cite{Frankfurt:1997zk}.
It was possible to extend this formalism by accounting for the well understood QCD phenomena to reconstruct the form of the 
distribution $P_{\gamma}(\sigma,W)$ ~\cite{Blaettel:1993ah,Frankfurt:1997zk}, where $W$ is the invariant photon--proton energy.
  While the form of $P_{\gamma}(\sigma,W)$ can be calculated
from the first principles only for small $\sigma$~\cite{Frankfurt:1996ri}, it can be constrained by the following
integral relations:
\begin{eqnarray}
\int d\sigma P_{\gamma}(\sigma,W) \sigma & \equiv & \langle \sigma \rangle =\sigma_{\gamma p}(W)\,, \nonumber\\ 
\int d\sigma P_{\gamma}(\sigma,W) \sigma^2 & \equiv & \langle \sigma^2 \rangle=16 \pi 
\frac{d \sigma_{\gamma p \to Xp}(W,t=0)}{dt} \,, 
\label{eq:P_gamma_const}
\end{eqnarray}
where $\sigma_{\gamma p}(W)$ is the total photon--nucleon cross section; $d \sigma_{\gamma p \to Xp}(W,t=0)/dt$ is the 
cross section of photon diffractive dissociation on the proton including the $\rho$ meson peak, which determines the 
dispersion of CFs  encoded in $P_{\gamma}(\sigma,W)$.  Note that the distribution $P_{\gamma}(\sigma,W)$ is not 
normalizable~\cite{Frankfurt:1996ri}, i.e., the integral  $\int d\sigma P_{\gamma}(\sigma,W)$ is 
divergent at the lower integration limit due to the infinite renormalization of the photon Green's function
(the vacuum polarization).

Therefore, to model CFs in the photon,  we build a model interpolating between 
the regimes of small and large $\sigma$. For the former, we use the color dipole model (CDM) of 
the photon wave function,  where the (usually virtual) photon is treated as superposition of quark--antiquark 
pairs (dipoles). The  dipoles interact with the target with cross sections given by the
factorization theorem of perturbative QCD  for small dipoles~\cite{Blaettel:1993rd}.  
Note that in the literature there is a popular  assumption 
that the contribution of light vector 
mesons to the photon--nucleon cross section is dual to the integral over the small 
 masses of $q\bar q$ pairs (for 
example, 
$M^2\le  \mbox{1 GeV}^2$ for $\rho, \omega$-mesons). 
The  CDM 
gives a reasonable description of CFs for  $\sigma \ll \sigma(\pi N)$. 
For large $\sigma$, $\sigma \gg \sigma(\pi N)$, the CFs are determined by 
non-perturbative effects both in terms of the photon configurations involved and the strength of the interaction.
Therefore, we use the modified VMD (mVMD) approach~\cite{Frankfurt:2015cwa} to model their effects. 

In our analysis we use the results of the approach
developed in~\cite{McDermott:1999fa}, which gives
 a  good description of the proton structure function $F_{2p}(x,Q^2)$ down to $Q^2\sim \mbox{0.3 GeV}^2$.
 In this approach, the dipole cross section $\sigma_{q \bar{q}}$ is built in a piece-wise form.
For small dipoles corresponding approximately to $d_t \leq 0.3-0.4$ fm, one has~\cite{Blaettel:1993rd}:
\begin{equation}
\sigma_{q \bar{q}}(W,d_t,m_q)=\frac{\pi^2}{3} d_t^2 \alpha_s(Q_{\rm eff}^2) x_{\rm eff} g(x_{\rm eff},Q_{\rm eff}^2) \,,
\label{eq:sigma_pQCD}
\end{equation}
where   $W$ is the invariant photon--nucleon center of mass energy, $Q_{\rm eff}^2=\lambda/d_t^2$ for light quarks and 
$Q_{\rm eff}^2=m_q^2+\lambda/d_t^2$ for heavy quarks; $x_{\rm eff}=4 m_q^2/W^2+0.75 \lambda/(W^2 d_t^2)$;
$m_q=300$ MeV for light $u$, $d$ and $s$ quarks and $m_c=1.5$ GeV.
This choice of the quark masses ensures that the average transverse size of $q\bar q$ configurations in the photon 
wave function is close to that of the pion, $d_{\pi}=0.65$ fm, and also leads to a smoother interpolation between small 
and large $\sigma$ regimes.  The parameter $\lambda=4$ is chosen to best reproduce the HERA data on diffractive $J/\psi$ 
photoproduction~\cite{Frankfurt:2000ez}.
Note, however, that heavy quarks give a very small contribution to the quantities we discuss below.

For large dipole sizes,  $\sigma_{q \bar{q}}$ is constrained to be equal to
the total pion-nucleon cross section at the appropriate energy at  $d_t =d_{\pi}=0.65$ fm
and to slowly grow for $d_t > 0.65$ fm. Finally, for the intermediate values of 
$0.3 - 0.4 < d_t < 0.65$ fm, $\sigma_{q \bar{q}}$ is modeled as a smooth interpolation
between the low-$\sigma_{q \bar{q}}$ (\ref{eq:sigma_pQCD}) and large-$\sigma_{q \bar{q}}$ limits.

As a result, one can write the interpolation formula for $\sigma_{\gamma p}(W)$ as
\begin{equation}
\sigma_{\gamma p}(W)=\sum_q e_q^2 \int dz \,d^2 d_t \sigma_{q \bar{q}}(W,d_t,m_q) 
|\Psi_{\gamma,T}(z,d_t,m_q)|^2  \,,
\label{eq:total_cs}
\end{equation}
where $z$ is the fraction of photon momentum carried by the quark in the dipole; $d_t$ is the transverse 
distance between  the  quark and the antiquark; $e_q$  are the quark charge.
The photon wave function squared in the mixed momentum--coordinate representation is given in~\cite{Nikolaev:1990ja}.

It is worth emphasizing here that the dominant contribution to $\sigma_{\gamma p}$ in Eq.~(\ref{eq:total_cs}) originates 
from the nonperturbative interactions of large-size multiparton hadron-like configurations in the photon wave function,
 which do not resemble  $q \bar q$  dipoles. Duality considerations suggest that the contribution of such configurations can be approximated using the lightest vector meson.  Hence, we first calculate $P_\gamma$ in the model of 
Eq.~(\ref{eq:total_cs}) and next match it at moderate $\sigma$ to the nonperturbative model for CFs for transitions to light mesons. 

\begin{figure}[th!]
\begin{center}
\epsfig{file=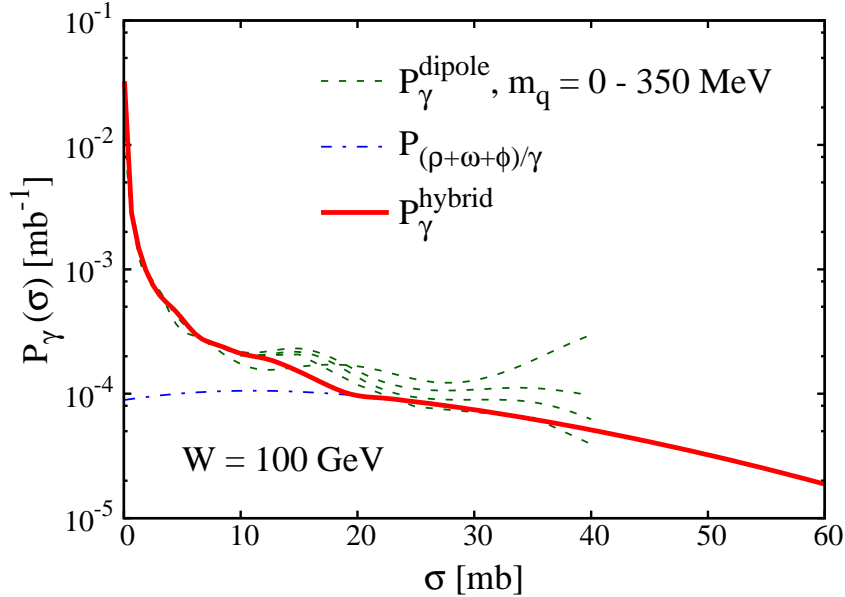,scale=0.9}
 \caption{The distributions $P_{\gamma}(\sigma,W)$ for the photon at $W=100$ GeV.
The red solid curve shows the  full result of the hybrid model, see Eq.~(\ref{eq:P_hybrid}).
The green dashed and blue dot-dashed curves show separately the dipole model and the vector meson contributions
evaluated using Eqs.~(\ref{eq:P}) and (\ref{eq:P_rho}),
respectively.
 }
 \label{fig:P_gamma}
\end{center}
\end{figure}

Since 
$\sigma_{\gamma p}(W)=\int d \sigma \sigma P_{\gamma}(\sigma,W) $, one finds within the model of
Eq.~(\ref{eq:total_cs}):
\begin{equation}
P_{\gamma}^{\rm dipole}(\sigma,W)=\sum_q e_q^2 \left|\frac{\pi dd_t^2}{d\sigma_{q \bar{q}}(W,d_t,m_q)}\right| 
\int dz |\Psi_{\gamma,T}(z,d_t(\sigma_{q \bar{q}}),m_q)|^{2}_{\ \big|\sigma_{q \bar{q}}(W,d_t,m_q)=\sigma}  \,.
\label{eq:P}
\end{equation}
Note that the right-hand side of (\ref{eq:P}) is expressed in terms of $\sigma_{q \bar{q}}(W,d_t,m_q)$, which is then
identified with $\sigma$.
The resulting distribution $P_{\gamma}^{\rm dipole}(\sigma,W)$ as a function of $\sigma$ for different light quark masses
$m_q$ and at $W=100$ GeV is  shown by the green dashed curves.
To examine the sensitivity  of $P_{\gamma}^{\rm dipole}(\sigma,W)$ to the choice $m_q$, we varied the light quark mass
in the interval $0 \leq m_q < 350$ MeV; the results are shown in Fig.~\ref{fig:P_gamma}, where the dashed curves from the upper to the lower one correspond to $m_q=0$, $m_q=250$ MeV, $m_q=300$, and  $m_q=350$ MeV,
respectively.

Since in the used 
model the $\sigma(q\bar q N)$ cross section
 does not exceed approximately 40 mb, the resulting 
distribution $P_{\gamma}^{\rm dipole}(\sigma,W)$ of Eq.~(\ref{eq:P}) has support only for $0 \leq \sigma \leq 40$ mb.

For large $\sigma$, the distribution $P_{\gamma}(\sigma,W)$ can be well approximated by 
the distribution $P(\sigma)$ for the $\gamma \to \rho$ transition.  Taking the sum 
of the $\rho$, $\omega$ and $\phi$ meson  contributions, the resulting distribution reads:
\begin{equation}
P_{(\rho+\omega+\phi)/\gamma}(\sigma,W)=\frac{11}{9}\left(\frac{e}{f_{\rho}}\right)^2 P(\sigma,W) \,,
\label{eq:P_rho}
\end{equation} 
where $P(\sigma,W)$ is taken from~\cite{Frankfurt:2015cwa}; the coefficient of $11/9$ takes into account the $\omega$ and 
$\phi$ contributions in the SU(3) approximation (which somewhat
overestimates the rather small contribution of $\phi$ mesons).
The form of $P(\sigma,W)$ is motivated by $P_{\pi}(\sigma,W)$ for
 the pion and takes into account presence of the large-mass diffraction at high energies.  
 It is also constrained to describe the 
 HERA data on $\rho$ photoproduction on the proton, which requires to account for a suppression  of the overlap of the 
 photon and $\rho$ wave function as compared to the diagonal case of the $\rho \to \rho$ transition.

 The resulting $P_{(\rho+\omega+\phi)/\gamma}(\sigma)$ at $W=100$ GeV 
 is shown by the blue dot-dashed curve in Fig.~\ref{fig:P_gamma}.

We build a hybrid model of $P_{\gamma}(\sigma,W)$ by interpolating between the regime
of small $\sigma \leq 10$ mb, where 
perturbative dipole approximation is applicable 
and there is no dependence on the light quark mass $m_q$,  and the regime of large $\sigma$,  
where the soft contribution due to the lightest vector meson dominates 
(hence we neglect the soft  contribution of configurations with the large mass and small $k_t$). 
In particular, in our analysis we use the following expression:
\begin{equation}
P_{\gamma}(\sigma,W)=\left\{\begin{array}{ll}
P_{\gamma}^{\rm dipole}(\sigma,W) \,, & \sigma \leq 10 \ {\rm mb} \,, \\
P_{\rm int}(\sigma,W) \,, & 10 \ {\rm mb} \leq \sigma \leq 20 \ {\rm mb} \,, \\
P_{(\rho+\omega+\phi)/\gamma}(\sigma,W) \,, &  \quad \sigma \geq 20 \ {\rm mb} \,.\end{array} \right. 
\label{eq:P_hybrid}
\end{equation}
where $P_{\rm int}(\sigma)$ is a smooth interpolating function.
The resulting $P_{\gamma}(\sigma,W)$ is shown by the red solid curve in Fig.~\ref{fig:P_gamma}. 

Our model for $P_{\gamma}(\sigma,W)$ satisfies the constraints of Eq.~(\ref{eq:P_gamma_const}) and gives the good 
description of the total and diffraction dissociation photon--proton cross sections at $W=100$ GeV.
Indeed, for $\sigma_{\gamma p}$, we obtain $\int_0^{100 \ {\rm mb}} d\sigma \sigma P_{\gamma}(\sigma,W)=135$ $\mu$b, 
which agrees with the PDG value of $\sigma_{\gamma p}=146$ $\mu$b~\cite{Agashe:2014kda}.
For the cross section of diffractive dissociation, we obtain
$\int_0^{100 \ {\rm mb}} d\sigma \sigma^2 P_{\gamma}(\sigma,W)/(16 \pi)=240$ $\mu$b/GeV$^2$. It agrees with our estimate 
of $d\sigma_{\gamma p \to Xp}(t=0)/dt \approx 220$ $\mu$b/GeV$^2$, which is obtained by
integrating the data of~\cite{Chapin:1985mf} over the produced diffractive masses and extrapolating the resulting
cross section to the desired $W=100$ GeV.

To quantify the width of CFs, one can introduce the dispersion $\omega_{\sigma}$. For the photon, it can be introduced by the following relation:
\begin{equation}
\int d\sigma \sigma^2 P_{\gamma}(\sigma,W)=(1+\omega_{\sigma}) \left(\frac{e}{f_{\rho}} \hat{\sigma}_{\rho N} \right)^2 \,,
\label{eq:omega_s}
\end{equation}
where $\hat{\sigma}_{\rho N}$ is the $\rho$ meson--nucleon cross section.
The use of our $P_{\gamma}(\sigma,W)$ in Eq.~(\ref{eq:omega_s}) gives $\omega_{\sigma} \approx 0.93$, which should be compared 
to $\omega_{\sigma}^{\rho} \approx 0.54$ for the pure $\rho$ meson contribution to $P_{\gamma}(\sigma,W)$ and to
$\omega_{\sigma}^{\pi} \approx 0.45$ for CFs in the pion~\cite{Blaettel:1993ah}.

\section{Color fluctuations and the number of wounded nucleons in $\gamma A$ scattering}
\label{sec:fluct}

One of important advantages of the Gribov--Glauber approximation  is that it accounts for  diffractive processes 
in the intermediate states  including the photon diffraction  into large masses and, therefore,  
conserves energy--momentum 
by virtue of duality between the parton model and hadronic descriptions.    
On the contrary, the Gribov--Glauber model,  which 
accounts for elastic intermediate state only~\cite{Bertocchi:1976bq}, violates energy--momentum conservation 
for the processes with multiple multiplicity of wound nucleons; 
it is proven by direct calculations of the energy released in such processes.
A Monte-Carlo procedure including finite size effects  in the elementary cross section and 
short-range correlations between nucleons was developed in~\cite{Alvioli:2013vk}.  
Thus, the formulae for the number of wounded nucleons follow directly 
from the formulae for the CFs but differ from  the combinatorics of the Glauber model due to the need to 
average over values of the cross section. For hard processes,  nuclear shadowing  and its impact on the number 
of wounded nucleons is calculated separately through the QCD factorization theorem. 
It has been understood long ago that the large coherence length 
prevents cascading of rapid secondary hadrons since they are formed outside of a target.  
Thus, only low-energy cascades are allowed.  Hence, the number of wounded nucleons given by the formulae below  
can be probed by selecting a kinematical region in the rapidity, where the contribution of 
cascades is expected to be small, see the discussion in the next section.

Previously we used the CF model to calculate the cross section of inelastic  interactions with exactly $\nu$ 
nucleons, $\sigma_\nu$, in $pA$ collisions. The model was found to be consistent with the data at least up 
to $\nu \sim 10$ \cite{Aad:2015zza}. Hence it is natural to use a similar approach to account for the   CF in 
the photon wave function in $\gamma A $ scattering for the interaction strength comparable or larger than 
$\sigma(\pi N)$ (CF effects due to   the contribution of small-size configurations to be discussed later, see
Eq.~(\ref{GGmod1})). Then, for the photon--nucleus cross  section corresponding to exactly $\nu$ inelastic 
interactions with the target nucleons, $\sigma_{\nu}$, one obtains in the 
Gribov--Glauber model  in the optical model limit:
\begin{equation}     
\sigma_{\nu} =  \int d\sigma 
    P_{\gamma}(\sigma,W) \left(\begin{array}{c}
      A\\
      \nu \end{array} \right)
     \int
    d^2\Vec{b}\, \left[{\sigma_{in}(\sigma) T_A(b)\over A}\right]^{\nu}
    \left[1-{\sigma_{in}(\sigma) T_A(b)\over A}\right]^{A-\nu} \,, 
  \label{GGg}
  \end{equation}
where $\vec{b}$ is the impact parameter;  $\sigma_{in}$ is the inelastic, non-diffractive cross section for 
the configuration characterized   by the total cross section $\sigma$;
 $T_A(b)=\int dz \rho_A(b,z)$ in the nuclear optical density, where $\rho_A(r)$ is the density of nucleons. 
Note that we use $\sigma_{in}=0.85\, \sigma$ (it is based on our estimate that in the considered range, 
the elastic cross section constitutes approximately 15\% of the total one)   and the Wood--Saxon density of 
nucleons for the $^{208}$Pb target~\cite{Alvioli:2013vk}   in our analysis. 
In the derivation of Eq.~(\ref{GGg}), we employ the discussed above equivalence between the Gribov--Glauber model 
and  cross section fluctuations  approach. This equivalence becomes trivial, if one uses the approximation of 
completeness over diffractively produced states.
It is worth emphasizing that we consider here soft interaction of the multiparton configurations of the
hadronic component of the photon wave function. For the interaction of the projectile consisting {\bf exactly} 
of two constituents, only $\nu =1, 2$ are allowed, see Ref.~\cite{Mandelstam:1963cw,Gribov:1968fc}.

\begin{figure}[h]
   \begin{center}
\epsfig{file=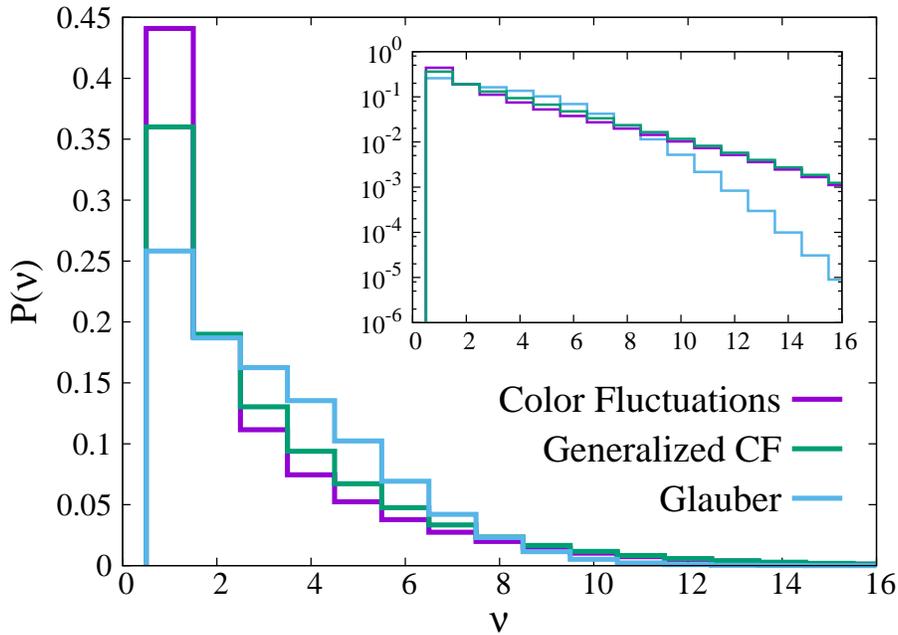,scale=0.95}
\caption{The probability distributions $P(\nu,W)$ of the number of inelastic collisions $\nu$.
Predictions of Eqs.~(\ref{GGg}) and (\ref{GGmod1}) are shown by the curves labeled ``Color Fluctuations''
and ``Generalized CF'', respectively.For comparison, the Gribov-Glauber model calculation with $\sigma=25$ mb, which neglects the effect of CFs,
is shown by the curve labeled ``Glauber''. }
     \label{fig:P_nu}
   \end{center}
 \end{figure}

The probability to have exactly $\nu $ wounded nucleons in $\gamma A$ scattering, $P(\nu)$, reads: 
\begin{equation}
P(\nu,W)= {\sigma_\nu\over \sum_1^\infty \sigma_\nu} \,,
\label{eq:p_nu}
\end{equation}
where $\sigma_{\nu}$ are given by Eq.~(\ref{GGg}). 
The  probability distribution $P(\nu,W)$ calculated using Eqs.~(\ref{GGg}) and (\ref{eq:p_nu}) is shown in
Fig.~\ref{fig:P_nu} by the curve labeled ``Color Fluctuations''. 
For comparison, we also show the results of the calculation, 
where the effect of CFs is neglected and the photon is represented by an effective fluctuation interacting with
the total cross section $\sigma=25$ mb; the corresponding curve is labeled ``Glauber''.

Equation~(\ref{GGg})  does not take into account that in QCD, configurations corresponding to a small cross 
section of the interaction with the nucleon at high energies 
interact with the collective small-$x $ gluon field of the nucleus, 
which is suppressed compared to the sum of the individual gluon fields of the nucleons due to the phenomenon 
of the  leading twist (LT) nuclear shadowing~\cite{Frankfurt:2011cs}.  This  is 
supported by  the observation of the large LT shadowing in  coherent photoproduction of $J/\psi$ in Pb-Pb UPCs at 
the LHC~\cite{Abelev:2012ba,Abbas:2013oua,Guzey:2013xba}.
This implies that Eq.~(\ref{GGg}) underestimates the probability of the interaction with two and more nucleons 
for small $\sigma$, which is determined by the LT nuclear shadowing.
 It effectively takes into account  the implication  of QCD factorization theorem: the presence of 
 the multiparton configurations in a small size $q\bar q$ configurations 
 which are ignored in the eikonal models and in particular in Eq.~(\ref{GGg}).

To take into the account 
this effect, we modify   Eq.~(\ref{GGg}) and use the following expression:
 \begin{eqnarray}
 \sigma_{\nu} =\int_0^\infty d\sigma 
   P_{\gamma}(\sigma,W) \left(\begin{array}{c}
      A\\
      \nu \end{array} \right) 
      \left[{\sigma^{in}\over \sigma_{\rm eff}^{in}} \Theta(\sigma_0-\sigma)+\Theta(\sigma-\sigma_0)\right]
    \int
    d^2\vec{b}\, \left[{\sigma_{\rm eff}^{in} T_A(b)\over A}\right]^{\nu}
    \left[1-{\sigma_{\rm eff}^{in} T_A(b)\over A}\right]^{A-\nu} \,,
      \label{GGmod1}
\end{eqnarray}
where  $\sigma_0=20$ mb (see details below); $\sigma^{in}/\sigma_{\rm eff}^{in} \approx \sigma/\sigma_{\rm eff} < 1$
 is the suppression factor  modeling the effect of the LT shadowing.
The effective cross section of $\sigma_{\rm eff}$ (note that $\sigma_{\rm eff}^{in}=0.85 \sigma_{\rm eff}$)
is a function of $\sigma$, which we determine using the following 
procedure. For large $\sigma > \sigma_0$, we set $\sigma_{\rm eff}=\sigma$. For $\sigma < \sigma_0$, 
$\sigma_{\rm eff}$ is defined as the cross section corresponding to the gluon shadowing ratio
$R_g(x)$~\cite{Frankfurt:2011cs} calculated in the high-energy eikonal approximation:
\begin{equation}
R_g(x_{\rm eff},Q^2_{\rm eff})=\frac{xg_A(x_{\rm eff},Q^2_{\rm eff})}{Axg_N(x_{\rm eff},Q^2_{\rm eff})}=\frac{2}{A \sigma_{\rm eff}} \int d^2 \vec{b} \left(1-e^{-\sigma_{\rm eff}/2 T_A(b)}\right) \,,
\label{eq:R_g}
\end{equation}
where $x_{\rm eff}$ and $Q^2_{\rm eff}$ are the light-cone momentum fraction and the resolution scale, respectively, which 
correspond to the dipole cross section for the given cross section 
$\sigma=\sigma_{q\bar{q}}(W,d_t,m_q)$ (the transverse size $d_t$), see Eq.~(\ref{eq:sigma_pQCD}).
This prescription for $\sigma_{\rm eff}$ is based on the observation that
since the non-vector-meson  component of  $P_\gamma (\sigma)$  is relatively small, 
the gluon shadowing can be considered in a simplified approximation, where CFs for the interaction with  
$N \ge 2 $ nucleons are small and, hence, $R_g$ is given by the single effective rescattering cross section
$\sigma_{\rm eff}$.

To estimate the value of $\sigma_0$, we notice that the factor of nuclear suppression of coherent $J/\psi$ 
photoproduction on nuclei is described very well for the LT nuclear shadowing. In particular, 
$R_g \approx 0.6$ for $x=10^{-3}$~\cite{Guzey:2013xba}, which 
 according to Eq.~(\ref{eq:R_g})   corresponds to
$\sigma_{\rm eff} =17$ mb. Therefore, in our analysis we take $\sigma_0=20$ mb. Our
 numerical analysis indicates that the results of our calculation depend weakly on the method of smooth interpolation 
in Eq.~(\ref{eq:P_hybrid})
and the assumption about  the value of the ratio 
$ \sigma^{in}/ \sigma^{in}_{\rm eff}$.  We call the resulting approach to the calculation of photon--nucleus 
inelastic cross sections $\sigma_{\nu}$ the generalized
color fluctuation (GCF) model.  The result of the calculation of the distribution over $\nu$ using Eq.~(\ref{GGmod1})
is shown in Fig.~\ref{fig:P_nu} by the curve labeled ``Generalized CF''.

The results presented in Fig.~\ref{fig:P_nu} deserve a discussion. For one inelastic photon--nucleus interaction
($\nu=1$), CFs in the photon lead to an almost a factor of two enhancement of $P(\nu)$ compared to the calculation neglecting CFs. Thus, an inclusion of the approximately 30\% small-$\sigma$ component of the photon wave function
(see the discussion in the Introduction), leads to a large effect in the inelastic $\gamma A$ scattering.
This effect is reduced approximately by a factor of two when we include the LT nuclear shadowing (compare 
the ``Color Fluctuations''
and ``Generalized CF'' curves). As $\nu$ increases, the small-$\sigma$ contribution to the distribution
$P_{\gamma}(\sigma,W)$ becomes progressively less important and all three models give similar results for $2 < \nu < 8$,
where the contribution of the two terms in the integrand of Eq.~(\ref{GGmod1}) approximately compensate each other.
For large $\nu > 10$, the two models including the effect of CFs in the photon predict a much broader distribution $P(\nu)$ 
than the model neglecting CFs: the enhancement at large $\nu$ comes from the contribution of the large-mass inelastic diffractive 
states implicitly included in Eqs.~(\ref{GGg}) and (\ref{GGmod1}).

\section{Color fluctuations and the distribution over transverse energy}
\label{sec:transverse_energy}

It is impossible to directly measure 
the number of inelastic interactions $\nu$ for collisions with nuclei. Modeling the distribution over the 
hadron multiplicity is also difficult due to the lack of the relevant data from  $\gamma p$ scattering and issues 
with implementing energy--momentum conservation. However,  the analysis of~\cite{Aad:2015zza} suggests that the 
distribution over the total transverse energy, $\Sigma{E}_T$, 
sufficiently far away from the projectile fragmentation region (at sufficiently large 
negative pseudorapidities) 
is  weakly  influenced by energy conservation effects
(due to the approximate Feynman scaling in this region) and is also weakly  
correlated with the activity in the rapidity-separated
forward region. This expectation is validated by a recent measurement
of $\Sigma{E}_T$ as a function of hard scattering kinematics in $pp$
collisions at the LHC~\cite{Aad:2015ziq}.

Due to the weak sensitivity to the projectile fragmentation region, we expect that the 
$\Sigma{E}_T$ distributions  in $pA$ and 
$\gamma A$ scattering  at similar energies should have similar shapes for the same $\nu$.
 In Ref.~\cite{Aad:2015zza}, a model was developed 
  for the distribution over $\Sigma{E}_T$ as a function of 
centrality in   $pA$ scattering at 
large negative pseudorapidities (in the Pb-going direction)
and   $\sqrt{s} = 5.02$ TeV. 
  In our discussion below, 
  using the one-to-one correspondence between centrality and $\nu$, we denote this distribution
 $f_{\nu}(\Sigma{E}_T)=1/N_{\rm evt}dN/d\Sigma{E}_T$. 
  In the spirit of the KNO scaling, it is natural to expect that the distribution over  
  the $\Sigma{E}_T$ total transverse energy in $\gamma A$ scattering, 
  when normalized to the average energy release in $pp$ scattering 
  $\left< \Sigma{E}_T (NN)\right>$, weakly depends on the incident collision energy.
 That is, the distribution over  $y = \Sigma{E}_T(\gamma N)/ \left< \Sigma{E}_T (\gamma N)\right>$ has approximately the same shape at 
 different energies. Hence we model the distribution over $y$ for photon--nucleus collisions using 
 $F_\nu(y) = \left< \Sigma{E}_T (NN)\right>f_\nu(y)$, where the factor of  $\left< \Sigma{E}_T (NN)\right>$ 
 is a Jacobian to keep normalization of $\int F_\nu(y) dy= P(\nu)$.

\begin{figure}[h]
\begin{center}
\epsfig{file=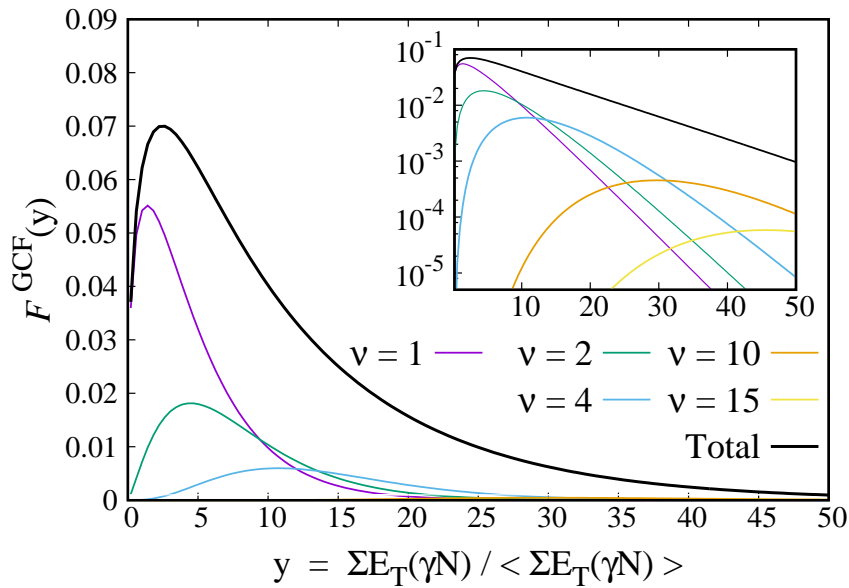,scale=0.9}
\caption{The probability distributions $F_\nu(y)$ over $y=\Sigma{E}_T(\gamma N) / \left< \Sigma{E}_T (\gamma N)\right> $ for different 
numbers of inelastic interactions $\nu$
in the Generalized Color Fluctuations (GCF) model.}
\label{fig:ET_1}
\end{center}
\end{figure}

 The results of the calculation   of $F_\nu(y)$  are presented in  Fig.~\ref{fig:ET_1}  for the Generalized Color 
 Fluctuations (GCF) model showing  contributions of events with different $\nu$ 
 to the normalized distribution over $y$. 
 We separately show the contributions corresponding to $\nu=1$, 2, 3, and 4, and the total contribution
 corresponding to the sum over all $\nu$ (the curve labeled ``Total'').  
  One can see that the  net distribution 
  is predicted to be much broader than that for the $\nu=1$ case corresponding to 
 the $\gamma p$ scattering. 
 Also, our results indicate that  for  $y=\Sigma{E}_T(\gamma N) / \left< \Sigma{E}_T (\gamma N)\right> \le 1 $,
 the contribution of the interactions with one nucleon dominates. 
 On the other hand, the distribution over 
 $y$  in $\gamma p$ scattering can be measured in $pA$  UPCs. 
 A first step would be to test that the $y$ distribution in $\gamma p$ and in the $\gamma A$ process with $\nu =1$ 
 [for example, in the interaction of the  
 direct photon ($x_{\gamma}=1$) 
  with a gluon  with  $x_A \ge 0.01$] is the same. 
  Among other things this would give a valuable information on the rapidity range affected by cascade interactions of 
  slow (in the nucleus rest frame) hadrons which maybe formed inside the nucleus.
  
\begin{figure}[h]
\begin{center}
\epsfig{file=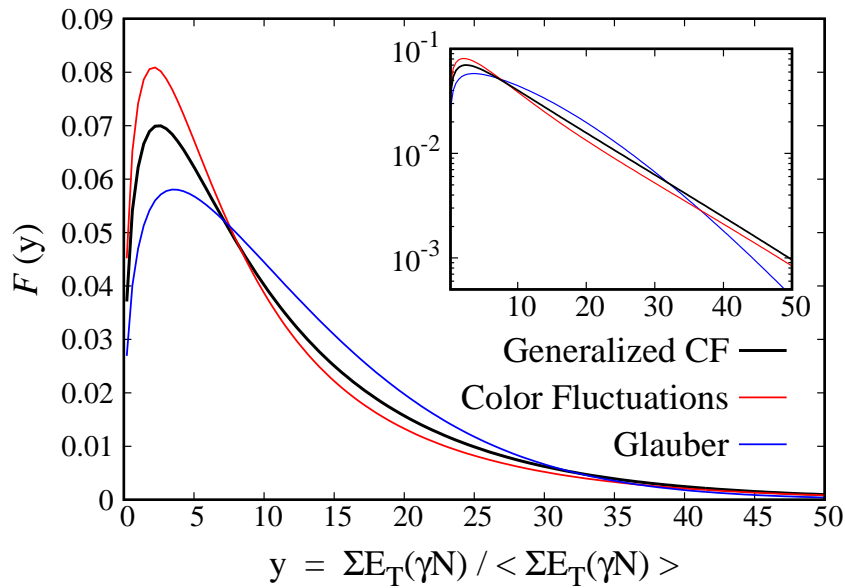,scale=0.9}
\caption{The net probability distribution $\sum _{\nu} F_\nu(y)$ as a function of $y$ for different models
including (curves labeled ``Generalized CFs'' and ``Color Fluctuations'') and neglecting (the curve labeled 
``Glauber'') CFs in the photon.
}
\label{fig:ET_2}
\end{center}
\end{figure}

  Next one would be able to  compare the rates of $y< 1$ events in $\gamma p$ and $\gamma A$ to
 determine the fraction of the $\nu =1$ and $\nu > 1$ events, which is quite sensitive to the model, 
 see Fig.~\ref{fig:P_nu}. 
One can see that for a given $y$, a range of $\nu$ contributes into the
cross section. To a good approximation,
$\left<\nu \right > - 1 \propto y$. For $y=10$,
 $\left<\nu \right > $ reaches 2.8 (2.6, 3.1) for the GCF
 (CF, Glauber) model with the variance typically of about
 $\sim 0.15$. The resulting  smearing  over $\nu$ for given $y$ does not
wipe out the difference between the models for the $\nu $ distribution,
see Fig.~\ref{fig:ET_2}.

Since  the distribution $F(y)$  is predicted to be much broader in $\gamma A$ collisions than in $\gamma p$ 
scattering, the use of different forward triggers makes it possible to determine the distribution over $\nu$ and 
use it to determine both $\left<\sigma\right >$ and the variance of the 
 $P_{\gamma}(\sigma,W)$ distribution for selected configuration. For example, 
in the CF model  of Eq.~(\ref{GGg})
({\it cf.}~\cite{Alvioli:2013vk,Alvioli:2014sba}), which does not include the LT shadowing effects, one 
obtains the following relations for the average number of inelastic collisions $\langle \nu \rangle$,
\begin{equation}
\left<\nu \right> = {A\sigma_{in}(\gamma N)\over \sigma_{in}(\gamma A)} \,,
\label{eq:nu_av}
\end{equation} and for the variance of the cross section for a specific trigger,
\begin{equation}
    {\left< \sigma_{trig}^2\right> \over \left<\sigma_{trig}\right>} = { (\left<\nu^2 \right>/\left<\nu \right> -1 )  
      {A^2\over A-1} \over \int d^2b \, T_A^2(b)} \,.
    \label{maver}
\end{equation}

Obviously similar considerations are applicable for the $\gamma A$ interactions with a special trigger including
 jet production, production of charm, etc. 
In the case of forward dijet production,
for direct photon for   
  $x_A \le 0.01$,  the  leading twist shadowing should set in resulting in a broader distribution over $\nu$ 
  as compared to the interactions with $x_A> 0.01$ (corresponding to $\nu=1$), see the 
  discussion in sections 6.3 and 6.4 of~\cite{Frankfurt:2011cs}.
For the resolved photons,   
the distribution over $\nu$ (and hence over $\Sigma E_T$) should 
become broader 
with an  decrease of  $x_\gamma$ 
since   hadronic configurations with smaller $x_\gamma$ have a larger transverse size.
One also expects that  for sufficiently small 
 $x_{\gamma} < 0.1$, the hard process would select generic configurations in the photon and, hence, the  distribution 
 over $\Sigma E_T$ would approach the distribution for  generic
 (without trigger)
  $\gamma A$ collisions. Note that first studies of diffractive dijet photoproduction in $pp$, $pA$ and $AA$ UPCs at the LHC 
in next-to-leading order (NLO) 
QCD, where CFs in the photon were used to model the effect of factorization breaking, were reported in~\cite{Guzey:2016tek}.

In the case of production of leading charm, small-size dipoles dominate 
(the variation of the transverse size is regulated by $m_c$ and $p_t(charm)$),
which allows one to study leading twist shadowing effects in the charm channel.
For instance, for $x\sim 10^{-3}$, one expects  $\left<\nu \right> \sim 2 $ and the corresponding reduction 
of $\sigma_{in}^{charm} (\gamma A)/A \sigma_{in}^{charm} (\gamma p)$, see Eq.~(\ref{eq:nu_av}).

\section{Conclusions}
\label{sec:conclusions} 

In this paper, we quantify the general property of photon--hadron interactions at high energies that the photon
can be viewed as a superposition of configurations interacting with different cross sections, which we call 
the phenomenon of color fluctuations (CFs), and propose a model for the distribution $P_{\gamma}(\sigma,W)$ describing these
CFs. Using this model and also additionally taking into account the effect of leading twist nuclear shadowing for
small-$\sigma$ configurations,  we for the first time give predictions for the distribution over the number of inelastic
interactions $\nu$ in photon--nucleus scattering. Our results show that  CFs lead to a dramatic enhancement of this distribution
at the small $\nu=1$ and the large $\nu > 10$ compared to the 
combinatorics familiar from the Glauber model. We also study the effect of CFs on the total transverse energy $\Sigma E_T$ released in inelastic $\gamma A$ 
scattering with different triggers and point to specific indications of the CF effect.
Our predictions can be tested in the photon--nucleus ($\gamma A$) interactions
in UPCs of ions at the LHC, which are characterized
by high-intensity fluxes of quasi-real photons in a wide energy spectrum and which can be viewed as
an effective ``strengthonometor'' of the different components of the photon wave function.

It would also allow one to obtain (using central tracking of the LHC detectors) unique information on the 
centrality dependence of the production of forward hadrons carrying a large fraction of the photon 
momentum ($x_F \ge 0.5$).  For soft interactions, experiments at fixed-target energies did indicate a 
strong suppression of the low-$p_t$  and large-$x_F$ hadron production. 
At the same time, very little experimental information 
is available on  suppression of the leading hadron production at the collider energies and on its $W$
dependence. These and other related topics will be discussed in more detail elsewhere.

An alternative approach of~\cite{Kancheli:1973vc} assumed the  dominance  of the single Pomeron exchange in the crossed 
channel, which branches into many Pomerons interacting with nucleons of nuclei.  
 This approach predicts absence  of the depletion of the yield
of leading hadrons in high energy hadron--nucleus collisions. Hence, it is in variance with the existing data, see e.g.  \cite{Adams:2006uz}.  The physics of the Pomeron branching may become important at the energies significantly exceeding energies achieved at UPCs  and at future $eA$ collider, which are the subject of this paper.

\acknowledgments

L.F.'s and M.S.'s research was supported by the US Department of Energy Office of Science, Office of Nuclear Physics under Award No. DE-FG02-93ER40771.


\begin{thebibliography}{99}

\bibitem{Gribov:1968ia}
  V.~N.~Gribov,
  Sov.\ J.\ Nucl.\ Phys.\  {\bf 9} (1969) 369
   [Yad.\ Fiz.\  {\bf 9} (1969) 640].
   
\bibitem{Gribov:1973jg}
  V.~N.~Gribov,
  hep-ph/0006158.
  
\bibitem{Bauer:1977iq} 
  T.~H.~Bauer, {\it et al.},
  Rev.\ Mod.\ Phys.\  {\bf 50}, 261 (1978)
  [Erratum-ibid.\  {\bf 51}, 407 (1979)].

\bibitem{Frankfurt:1994hf} 
  L.~L.~Frankfurt, G.~A.~Miller and M.~Strikman,
  Ann.\ Rev.\ Nucl.\ Part.\ Sci.\  {\bf 44}, 501 (1994)
  [hep-ph/9407274].

\bibitem{Frankfurt:2000tya} 
  L.~Frankfurt, V.~Guzey and M.~Strikman,
  J.\ Phys.\ G {\bf 27}, R23 (2001)
  [hep-ph/0010248].


\bibitem{Dutta:2012ii} 
  D.~Dutta, K.~Hafidi and M.~Strikman,
  Prog.\ Part.\ Nucl.\ Phys.\  {\bf 69}, 1 (2013)
  [arXiv:1211.2826 [nucl-th]].

\bibitem{Mandelstam:1963cw} 
  S.~Mandelstam,
  Nuovo Cim.\  {\bf 30}, 1148 (1963).

\bibitem{Gribov:1968jf} 
  V.~N.~Gribov,
  Sov.\ Phys.\ JETP {\bf 29}, 483 (1969)
  [Zh.\ Eksp.\ Teor.\ Fiz.\  {\bf 56}, 892 (1969)].

\bibitem{Alvioli:2014sba} 
  M.~Alvioli, L.~Frankfurt, V.~Guzey and M.~Strikman,
  Phys.\ Rev.\ C {\bf 90}, 034914 (2014)
  [arXiv:1402.2868 [hep-ph]].

\bibitem{Alberi:1981af}
  G.~Alberi and G.~Goggi,
  Phys.\ Rept.\  {\bf 74} (1981) 1.

\bibitem{Good:1960ba} 
  M.~L.~Good and W.~D.~Walker,
  Phys.\ Rev.\  {\bf 120}, 1857 (1960).

\bibitem{Miettinen:1978jb} 
  H.~I.~Miettinen and J.~Pumplin,
  Phys.\ Rev.\ D {\bf 18}, 1696 (1978).

 \bibitem{Frankfurt:1997zk} 
  L.~Frankfurt, V.~Guzey and M.~Strikman,
  Phys.\ Rev.\ D {\bf 58}, 094039 (1998)
  [hep-ph/9712339].

\bibitem{Baym:1995cz} 
  G.~Baym, B.~Blattel, L.~L.~Frankfurt, H.~Heiselberg and M.~Strikman,
  Phys.\ Rev.\ Lett.\  {\bf 67} (1991) 2946; 
  Phys.\ Rev.\ C {\bf 52}, 1604 (1995)
  [nucl-th/9502038].



\bibitem{Aad:2015zza} 
  G.~Aad {\it et al.} [ATLAS Collaboration],
  Eur.\ Phys.\ J.\ C {\bf 76}, no. 4, 199 (2016)
  [arXiv:1508.00848 [hep-ex]].
 
 
 
 \bibitem{Bertocchi:1976bq} 
  L.~Bertocchi and D.~Treleani,
  J.\ Phys.\ G {\bf 3}, 147 (1977).
  
\bibitem{Guzey:2005tk} 
  V.~Guzey and M.~Strikman,
  Phys.\ Lett.\ B {\bf 633}, 245 (2006)
  [Phys.\ Lett.\ B {\bf 663}, 456 (2008)]
  [hep-ph/0505088].
 
\bibitem{Alvioli:2013vk} 
  M.~Alvioli and M.~Strikman,
  Phys.\ Lett.\ B {\bf 722}, 347 (2013)
  [arXiv:1301.0728 [hep-ph]].
  
   
  \bibitem{ATLAS:2014cpa} 
  G.~Aad {\it et al.} [ATLAS Collaboration],
  Phys.\ Lett.\ B {\bf 748}, 392 (2015)
    [arXiv:1412.4092 [hep-ex]].
  %
  %
  
\bibitem{Adare:2015gla} 
  A.~Adare {\it et al.} [PHENIX Collaboration],
  Phys.\ Rev.\ Lett.\  {\bf 116}, no. 12, 122301 (2016)
  [arXiv:1509.04657 [nucl-ex]].
   
  \bibitem{Alvioli:2014eda} 
  M.~Alvioli, B.~A.~Cole, L.~Frankfurt, D.~V.~Perepelitsa and M.~Strikman,
  Phys.\ Rev.\ C {\bf 93}, no. 1, 011902 (2016)
  [arXiv:1409.7381 [hep-ph]].
  
  \bibitem{Adam:2015gsa} 
  J.~Adam {\it et al.} [ALICE Collaboration],
  JHEP {\bf 1509}, 095 (2015)
  [arXiv:1503.09177 [nucl-ex]].
  
\bibitem{Frankfurt:2015cwa} 
  L.~Frankfurt, V.~Guzey, M.~Strikman and M.~Zhalov,
  Phys.\ Lett.\ B {\bf 752}, 51 (2016)
  [arXiv:1506.07150 [hep-ph]].
  
 


\bibitem{Baltz:2007kq} 
  A.~J.~Baltz {\it et al.},
  Phys.\ Rept.\  {\bf 458}, 1 (2008)
  [arXiv:0706.3356 [nucl-ex]].
  
\bibitem{Accardi:2012qut} 
  A.~Accardi {\it et al.},
  arXiv:1212.1701 [nucl-ex].

\bibitem{Boer:2011fh} 
  D.~Boer {\it et al.},
  arXiv:1108.1713 [nucl-th].
  \bibitem{Altarelli:2001ji}
  G.~Altarelli, R.~D.~Ball and S.~Forte,
  Nucl.\ Phys.\ B {\bf 621} (2002) 359
  [hep-ph/0109178].
  
  
\bibitem{Guzey:2013jaa} 
  V.~Guzey, M.~Strikman and M.~Zhalov,
  Eur.\ Phys.\ J.\ C {\bf 74}, no. 7, 2942 (2014)
  [arXiv:1312.6486 [hep-ph]].
   
  \bibitem{Gribov:1968gs}
  V.~N.~Gribov,
  Sov.\ Phys.\ JETP {\bf 30}, 709 (1970)
  [Zh.\ Eksp.\ Teor.\ Fiz.\  {\bf 57}, 1306 (1969)].
 
\bibitem{Sakurai:1960ju} 
  J.~J.~Sakurai,
  Annals Phys.\  {\bf 11}, 1 (1960).

\bibitem{Feynman:1973xc} 
  R.~P.~Feynman, {\it Photon-hadron interactions}, 
  (Benjamin, Reading 1972), 282 p.
 
 \bibitem{Bjorken:1972uk} 
  J.~D.~Bjorken,
  Conf.\ Proc.\ C {\bf 710823}, 281 (1971).
  
\bibitem{Frankfurt:1988nt} 
  L.~L.~Frankfurt and M.~I.~Strikman,
  Phys.\ Rept.\  {\bf 160}, 235 (1988).
  

\bibitem{Kopeliovich:1978qz}
  B.~Z.~Kopeliovich and L.~I.~Lapidus,
  Pisma Zh.\ Eksp.\ Teor.\ Fiz.\  {\bf 28} (1978) 664.
    
  \bibitem{Blaettel:1993ah} 
  B.~Blaettel, {\it et al.},
  Phys.\ Rev.\ D {\bf 47}, 2761 (1993).
 
 
\bibitem{Frankfurt:1996ri} 
  L.~Frankfurt, A.~Radyushkin and M.~Strikman,
  Phys.\ Rev.\ D {\bf 55}, 98 (1997)
  [hep-ph/9610274].
 
 
  \bibitem{McDermott:1999fa} 
  M.~McDermott, L.~Frankfurt, V.~Guzey and M.~Strikman,
  Eur.\ Phys.\ J.\ C {\bf 16}, 641 (2000)
  [hep-ph/9912547]. 
 
  \bibitem{Blaettel:1993rd} 
  B.~Blaettel, G.~Baym, L.~L.~Frankfurt and M.~Strikman,
  Phys.\ Rev.\ Lett.\  {\bf 70}, 896 (1993);
  L.~Frankfurt, G.~A.~Miller and M.~Strikman,
  Phys.\ Lett.\ B {\bf 304}, 1 (1993).

 \bibitem{Frankfurt:2000ez} 
  L.~Frankfurt, M.~McDermott and M.~Strikman,
  JHEP {\bf 0103}, 045 (2001)
  [hep-ph/0009086].

  
  \bibitem{Nikolaev:1990ja} 
  N.~N.~Nikolaev and B.~G.~Zakharov,
  Z.\ Phys.\ C {\bf 49}, 607 (1991).
 
  
\bibitem{Agashe:2014kda} 
  K.~A.~Olive {\it et al.} [Particle Data Group Collaboration],
  Chin.\ Phys.\ C {\bf 38}, 090001 (2014).
  

\bibitem{Chapin:1985mf} 
  T.~J.~Chapin {\it et al.},
  Phys.\ Rev.\ D {\bf 31}, 17 (1985).

\bibitem{Gribov:1968fc}
  V.~N.~Gribov,
  Sov.\ Phys.\ JETP {\bf 26} (1968) 414
   [Zh.\ Eksp.\ Teor.\ Fiz.\  {\bf 53} (1967) 654].

  \bibitem{Frankfurt:2011cs} 
  L.~Frankfurt, V.~Guzey and M.~Strikman,
  Phys.\ Rept.\  {\bf 512}, 255 (2012)
  [arXiv:1106.2091 [hep-ph]].
  
  \bibitem{Abelev:2012ba} 
  B.~Abelev {\it et al.} [ALICE Collaboration],
  Phys.\ Lett.\ B {\bf 718}, 1273 (2013)
  [arXiv:1209.3715 [nucl-ex]].
  
\bibitem{Abbas:2013oua} 
  E.~Abbas {\it et al.} [ALICE Collaboration],
  Eur.\ Phys.\ J.\ C {\bf 73}, no. 11, 2617 (2013)
  [arXiv:1305.1467 [nucl-ex]].


\bibitem{Guzey:2013xba} 
  V.~Guzey, E.~Kryshen, M.~Strikman and M.~Zhalov,
  Phys.\ Lett.\ B {\bf 726}, 290 (2013)
  [arXiv:1305.1724 [hep-ph]].


\bibitem{Aad:2015ziq} 
  G.~Aad {\it et al.} [ATLAS Collaboration],
  Phys.\ Lett.\ B {\bf 756}, 10 (2016)
  [arXiv:1512.00197 [hep-ex]].
  
  

\bibitem{Guzey:2016tek} 
  V.~Guzey and M.~Klasen,
  JHEP {\bf 1604}, 158 (2016)
  [arXiv:1603.06055 [hep-ph]].

    
  
   
\bibitem{Kancheli:1973vc}
  O.~V.~Kancheli,
  Pisma Zh.\ Eksp.\ Teor.\ Fiz.\  {\bf 18} (1973) 465
   [JETP Lett.\  {\bf 18} (1973) 274].
 
 \bibitem{Adams:2006uz}
  J.~Adams {\it et al.} [STAR Collaboration],
  Phys.\ Rev.\ Lett.\  {\bf 97} (2006) 152302,  [arXiv:nucl-ex/0602011].


\end{thebibliography}
\end{document}